\documentclass[aps,prb,preprint,floatfix,longbibliography]{revtex4-2}

\newcommand {\EF}{$E_{\rm F}$}
\newcommand {\vF}{$v_{\rm{F}}$}

\newcommand {\kF}{$k_{\rm{F}}$}
\newcommand {\BF}{$B_{\rm{F}}$}
\newcommand {\tauQ}{$\tau_{\rm Q}$}
\newcommand {\tautr}{$\tau_{\rm tr}$}

\newcommand {\Jeff}{\it J$_{\rm{eff}}$}

\newcommand {\AIrO}{$A$IrO$_3$}
\newcommand {\CaIrO}{CaIrO$_3$}

\newcommand {\CdAs}{Cd$_3$As$_2$}

\newcommand {\CaCO}{CaCO$_3$}
\newcommand {\IrO}{IrO$_2$}
\newcommand {\Rhoxx}{$\rho_{xx}$}
\newcommand {\Rhoyx}{$\rho_{yx}$}
\newcommand {\RhoxxFit}{$\rho_{xx}^{\rm fit}$}
\newcommand {\DeltaRhoxx}{$\Delta \rho_{xx}$}
\newcommand {\Gxy}{$\sigma_{xy}$}

\newcommand {\nH}{$n_{\rm H}$}

\newcommand {\mutr}{$\mu _{\rm tr}$}

\newcommand {\nUnit}{$\rm{cm}^{-3}$}
\newcommand {\muUnit}{$\rm{cm}^{2}/\rm{Vs}$}

\newcommand {\nthreeD}{$n _{\rm 3D}$}
\newcommand {\Bparaa}{$B \parallel a$}
\newcommand {\Bparac}{$B \parallel c$}

\newcommand {\mass}{$m_{\rm c} / m_{0}$}
\newcommand {\massc}{$m_{\rm c}$}
\newcommand {\massz}{$m_{0}$}

\usepackage{graphicx}
\usepackage{epstopdf}

\usepackage{color}
\usepackage{amsmath,bm}
\usepackage[utf8]{inputenc}
\usepackage{xcolor}
\usepackage{physics}
\usepackage{mathcomp}
\usepackage{comment}
\usepackage[resetlabels,labeled]{multibib}
\usepackage[normalem]{ulem}
\usepackage{cancel}

\usepackage{xr} 
\begin{document}

\title[]{Carrier-density dependence of magnetotransport\\ in correlated Dirac semimetal {\CaIrO}}


\author{Rinsuke Yamada$^{1}$}\email{ryamada@ap.t.u-tokyo.ac.jp}
\author{Jun Fujioka$^{2,3}$}\email{fujioka@ims.tsukuba.ac.jp}
\author{Minoru Kawamura$^{4}$}
\author{Tatsuya Okawa$^{1}$}
\author{Yoshio Kaneko$^{4}$}
\author{Shiro Sakai$^{5}$}
\author{Motoaki Hirayama$^{4}$}
\author{Ryotaro Arita$^{4.6}$}
\author{Kiyohiro Adachi$^{4}$}
\author{Daisuke Hashizume$^{4}$}
\author{Yoshinori Tokura$^{1,4,7}$}\email{tokura@riken.jp}

\affiliation{$^{1}$Department of Applied Physics, The University of Tokyo, Bunkyo-ku, Tokyo 113-8656, Japan} 
\affiliation{$^{2}$ Department of Materials Science, University of Tsukuba, Tsukuba, Ibaraki 305-8573, Japan}
\affiliation{$^{3}$ Research Center for Organic-Inorganic Quantum Spin Science and Technology (OIQSST), University of Tsukuba,
Tsukuba, Ibaraki 305-8573, Japan}
\affiliation{$^{4}$ RIKEN Center for Emergent Matter Science (CEMS), Wako 351-0198, Japan}
\affiliation{$^{5}$Physics Division, Sophia University, Chiyoda-ku, Tokyo 102-8554, Japan} 
\affiliation{$^{6}$Department of Physics, The University of Tokyo, Bunkyo-ku, Tokyo 113-8656, Japan} 
\affiliation{$^{7}$Tokyo College, The University of Tokyo, Bunkyo-ku, Tokyo 113-8656, Japan}

\maketitle


\begin{center}
\Large{Abstract} 
\end{center}

\textbf{We report the carrier density dependence of the magnetotransport property in the correlated Dirac semimetal {\CaIrO}. In the dilute carrier density region ({\nH} $\sim 2.2 \times 10^{16} \,${\nUnit}) at $2 \, \mathrm{K}$, the mobility exceeds $1.0 \times 10^{5} \,${\muUnit} at $2 \, \mathrm{K}$, and the transverse magnetoresistance (MR) reaches $2,000 \,$\% at $12 \, \mathrm{T}$. The analysis of quantum oscillations and Hall conductivity shows that the Fermi velocity is nearly independent of the cross-sectional area of the Fermi surface, or equivalently the carrier density, supporting a $k$-linear dispersion of the Dirac node. The field dependence of magnetoresistivity is nearly $B$-linear in the moderate carrier density region ($n_\mathrm{H} \geq 4 \times 10^{16}\,$cm$^{-3}$), but scales with $B^{\alpha}$ ($\alpha > 2$) in the lower carrier density region. The variation of magnetoresistivity is likely affected by the enhanced long-range Coulomb interaction in the quantum limit, where Dirac electrons are subject to the magnetic confinement.
}

\newpage

\begin{center}
\Large{Main text}
\end{center}

\section{Introduction}
The magnetotransport of Dirac/Weyl electrons in topological semimetals has been intensively studied in pursuit of emergent topological phenomena. A striking feature of Dirac or Weyl electrons is the ultrahigh mobility, which is typically seen in magnetotransport phenomena such as quantum Hall effect and giant magnetoresistance (MR) \cite{2018ArmitageReview, 2005YZhangNature, 2017MUchidaNatCommun, 2018MGoyalAPLMater, 1953PBAlersPhysRev, 2014MNAliNature, 2015JXiongSience, 2014TLiangNatMater, 2016SciAdvHMatsuda, 2021JPSJDHirai}. For example, the electron mobility reaches the order of $10^7$ {\muUnit} and the giant positive MR with MR ratio of more than 100,000 \% is observed in the typical Dirac semimetal of {\CdAs} \cite{2014TLiangNatMater}. In particular, the transverse MR is enhanced in the dilute carrier density and high mobility regime, the mechanism of which has been argued in terms of field-induced variation of electronic states or scattering from lattice inhomogeneity/defects. In general, such magnetotransport phenomena have been intensively studied in topological semimetals with weak electron correlation effect, but recent research has uncovered that the quantum transport of high mobility Dirac electrons can be observed even in the correlated electron system \cite{2019FujiokaNatCommun}. In particular, it is demonstrated that the perovskite \CaIrO{} exhibits the Dirac semimetallic state with dilute carrier density and high mobility even in the strong electron correlation regime near the Mott criticality. Such a strongly correlated Dirac electron may show transverse MR  different from that in a conventional topological semimetal, yet they have rarely been explored experimentally \cite{2015KUedaPRL, 2015ZTianNatPhys, 2016DLiuPRL, 2018CYGuoNatCommun, 2020NatCommunKTakiguchi, 2021SciAdvJMOk}.

The perovskite {\AIrO} ($A$ = Ca, Sr) is a correlated Dirac semimetal, which offers a fertile ground to study the transport of highly mobile Dirac electrons near a Mott criticality. {\AIrO} crystallizes in an orthorhombic perovskite structure with GdFeO$_3$-type lattice distortion as shown in Fig. \ref{Structure_Linenode}(a). The electronic state near the Fermi energy ({\EF}) is mainly composed of a nearly half-filled {\Jeff} $=1/2$ multiplet of 5$d$ orbitals of the Ir$^{4+}$ ions, leading to a semimetallic state possessing electron pockets with Dirac dispersion and hole pocket(s) with a relatively flat band due to the interplay between the strong spin-orbit coupling and electron correlation \cite{2012JMCarterPRB, 2012MAZebPRB, 2013HZhangPRL, 2017FujiokaPRB, 2015NiePRL}. The theoretical calculations predict that the Dirac line node emerges around the U-point in the momentum space, which is protected by the nonsymmorphic crystalline symmetry and time reversal symmetry~\cite{2012JMCarterPRB}. The schematic illustration of the line node around the U-point is shown in Fig. \ref{Structure_Linenode}(b). On the basis of the transport measurements and theoretical calculations, it is revealed that the semimetallic state with dilute carrier density ($2 \times 10^{17} \, \mathrm{cm}^{-3}$) and high electron mobility ($> 6 \times 10^4 \, \mathrm{cm}^2/\mathrm{Vs}$) emerges at low temperatures, wherein the Dirac line node is closely located to {\EF} ($\sim$10 meV below {\EF}) for single-crystalline {\CaIrO}~\cite{2019FujiokaNatCommun, 2019RYamadaPRL, 2019MMasukoAPL, 2021JFujiokaPRB, 2022RYamadanpj, 1998ImadaRevModPhys}.
In particular, high-mobility Dirac electrons show giant longitudinal- and transverse-MR in the quantum limit (QL), where a quasi-one-dimensional (quasi-1D) state is realized along the field direction. 
On the basis of transport measurements in a longitudinal configuration, i.e., the conduction along the quasi-1D state, the possibility of a disordered charge density wave state is proposed \cite{2022RYamadanpj}. Nevertheless, the origin of the giant positive MR in the transverse configuration, i.e. the conduction in the direction perpendicular to the quasi-1D state, as well as the mechanism behind the high mobility, has been poorly understood. 

In this study, we have investigated the carrier density dependence of the magnetotransport property of {\CaIrO} to reveal the origin of the transverse MR emerging in the vicinity of the Mott transition. The mobility and MR ratio increase with decreasing the carrier density {\nH} and exceed $1.0 \times 10^5 \,${\muUnit} and $2,000 \,$\% (at $12 \, \mathrm{T}$) in the dilute carrier density region of {\nH} $\sim 2.2 \times 10^{16} \,${\nUnit} at $2 \, \mathrm{K}$. From the analysis of the Shubnikov-de Haas (SdH) oscillations, we demonstrate that the effective mass decreases with decreasing carrier density due to the nearly $k$-linear band dispersion. Furthermore, the MR shows $B$-nonlinear dependence with the exponent exceeding 2 in the low-{\nH} regime, which is likely induced by the enhanced long-range Coulomb interaction.

\section{Experimental methods}
The single-crystalline perovskite {\CaIrO} is prepared using a cubic anvil-type facility \cite{2019FujiokaNatCommun}. A stoichiometric mixture of high-purity {\CaCO} and {\IrO} powders was heated at 1000$\,{}^\circ\mathrm{C}$ for $24 \, \mathrm{h}$ at ambient pressure before the high-pressure synthesis. The synthesis temperature and pressure are $1200 \, {}^\circ\mathrm{C}$ and $1 \, \mathrm{GPa}$, respectively. A typical size of the single crystalline sample is about $0.4 \times 0.3 \times 0.2 \, \mathrm{mm}$. The crystal orientation was determined by the X-ray diffraction. The resistivity ({\Rhoxx}) and the Hall resistivity ({\Rhoyx}) are measured by a standard four-probe technique from $2 \, \mathrm{K}$ to $300 \, \mathrm{K}$ and up to $14 \, \mathrm{T}$. The transverse MR is measured under the magnetic field $B$ along the $c$-axis with the electric current $I$ along the $a$-axis or $b$-axis. On the contrary, the longitudinal MR is measured for $B \parallel I \parallel a$.

\section{Results and Discussion}
\subsection{Large transverse magnetoresistance.}
\begin{figure}[h]%
\centering
\includegraphics[trim=0cm 0.1cm 0.0cm 0.0cm,clip,width=0.95\textwidth]{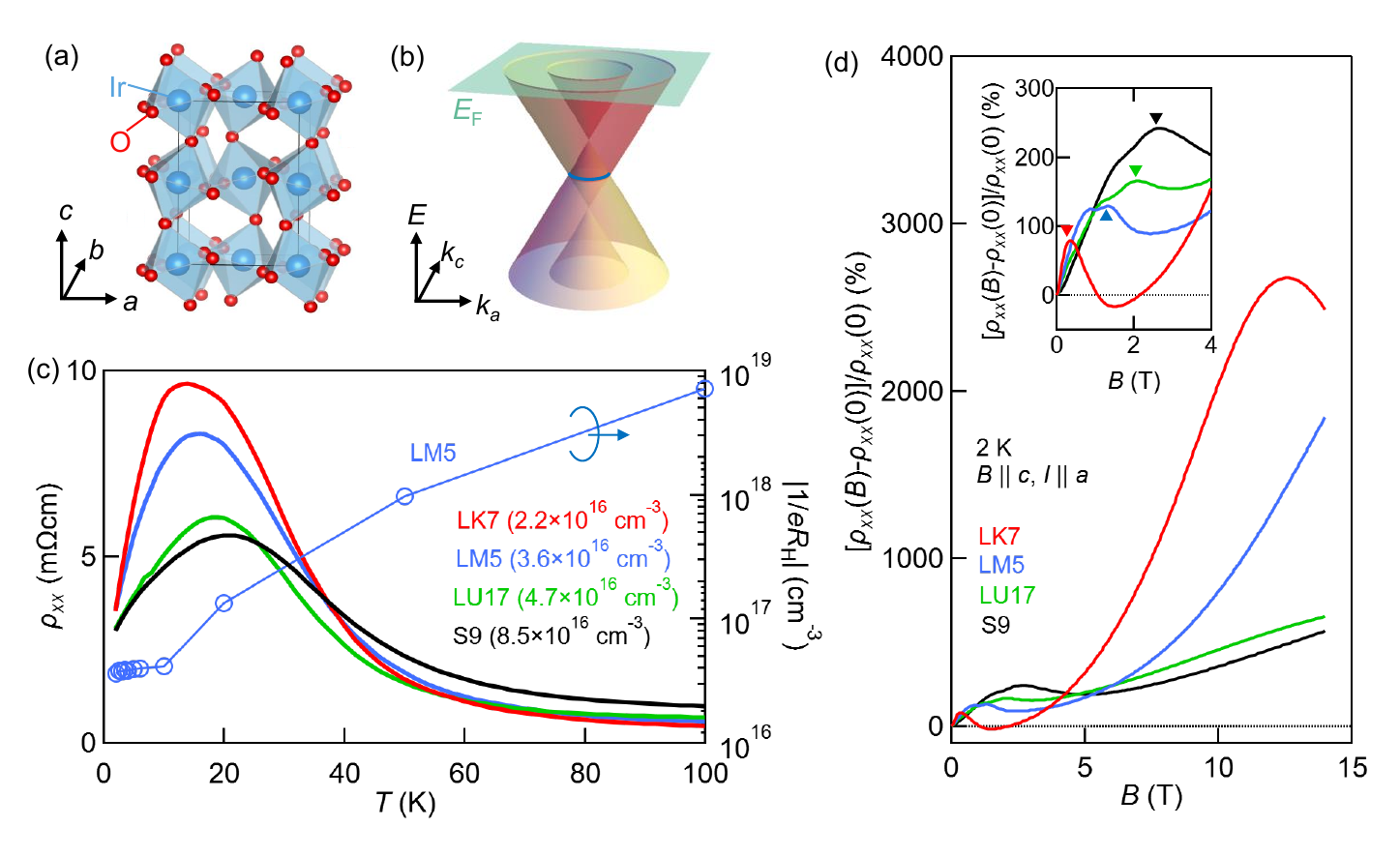}
\caption{\textbf{Filling dependence of basic transport responses.} 
(a) Crystal structure of orthorhombic perovskite {\CaIrO}. 
 (b) The schematic illustration of energy dispersion in the $k_a$$k_c$- plane near the line node (blue line). {\EF} denotes the position of the Fermi energy.
 (c) Temperature dependence of the resistivity of typical three samples with different carrier densities (left axis) and that of the carrier density {\nH} estimated from the Hall coefficient of sample LM5 (right axis).
(d) Transverse magnetoresistivity (MR) ratio defined by $[${\Rhoxx}$(B)-${\Rhoxx}$(0)]/${\Rhoxx}$(0)$ of typical four samples at 2 K. The inset shows the magnified view of the low-field region. Triangles in the inset indicate the magnetic field where SdH oscillations for {\Bparac} reach the quantum limit. Here, we applied the electric current along $a$ axis and the magnetic field along $c$ axis of the crystal. 
}\label{Structure_Linenode}
\end{figure}

First, we compare the basic transport properties of four typical samples of {\CaIrO} with a different band filling in Fig. \ref{Structure_Linenode}. The estimated carrier densities of sample LK7, LM5, LU17, and S9 from the Hall effect at $2 \, \mathrm{K}$ ($n_\mathrm{H}$) are 2.2, 3.6, 4.7, and 8.5 $\times 10^{16} \,${\nUnit}, respectively [see Table \ref{Table_SampleList}]. The variation in $n_\mathrm{H}$ likely originates from the slight difference in chemical composition, which is, however, too small to be reliably characterized by a standard chemical composition analysis (less than $0.001 \, \%$ of the density of Ir atoms). 
The resistivity curves in these four samples commonly show a peak around $20 \, \mathrm{K}$, and the peak position shifts to lower temperatures with decreasing {\nH} ($=|1/eR_\mathrm{H}|$) [see Fig. \ref{Structure_Linenode}(c)]. We estimate the Hall coefficient $R_\mathrm{H}$ using the slope of the Hall resistivity at zero field. {\nH} decreases with lowering temperature due to the reduction of thermally excited carriers and becomes almost independent of temperature below $10 \, \mathrm{K}$. Since the carrier mobility monotonically increases with lowering temperatures~\cite{2019FujiokaNatCommun, 2019RYamadaPRL}, a balance between the carrier density and mobility likely causes the peak of the resistivity around $20 \, \mathrm{K}$. The density of thermally excited carriers is sufficiently suppressed below $20 \, \mathrm{K}$, validating the use of a single-carrier model within semiclassical transport theory at low temperature and low magnetic field~\cite{2019FujiokaNatCommun}. Within this framework, the shift in the resistivity peak can be consistently attributed to the suppressed contribution of thermal activation with increasing {\nH}.

The field dependence of MR shows qualitatively different behaviors among the four samples. Figure \ref{Structure_Linenode}(d) shows the MR ratio at $2 \, \mathrm{K}$. The MR of S9, which has the highest carrier density among the four samples, has a peak at $2.5 \, \mathrm{T}$ and nearly linearly increases as a function of $B$ above $4 \, \mathrm{T}$. On the other hand, the MRs of LM5 and LK7 (see Table~\ref{Table_SampleList}) with relatively low carrier densities show a peak around $0.5 \sim 1.5 \, \mathrm{T}$, and then increase nonlinearly as a function of $B$. Furthermore, the MR of LK7 shows a downturn above $12 \, \mathrm{T}$. Overall, in samples with lower carrier density, the peak between $0.5$ and $3 \, \mathrm{T}$ shifts to lower magnetic fields, and MR ratio $= [\rho_{xx}(B)-\rho_{xx}(0)]/\rho_{xx}(0)$ becomes larger at a high magnetic field. At $2 \, \mathrm{K}$ and $12 \, \mathrm{T}$, the MR ratio of S9, LU17 and LM5 are about $460 \, \%$, $560 \, \%$, and $1,280 \, \%$, respectively, and that of LK7 reaches $2,640 \, \%$.

\begin{figure}[h]%
\centering
\includegraphics[trim=0cm 0.1cm 0.0cm 0.0cm,clip,width=0.95\textwidth]{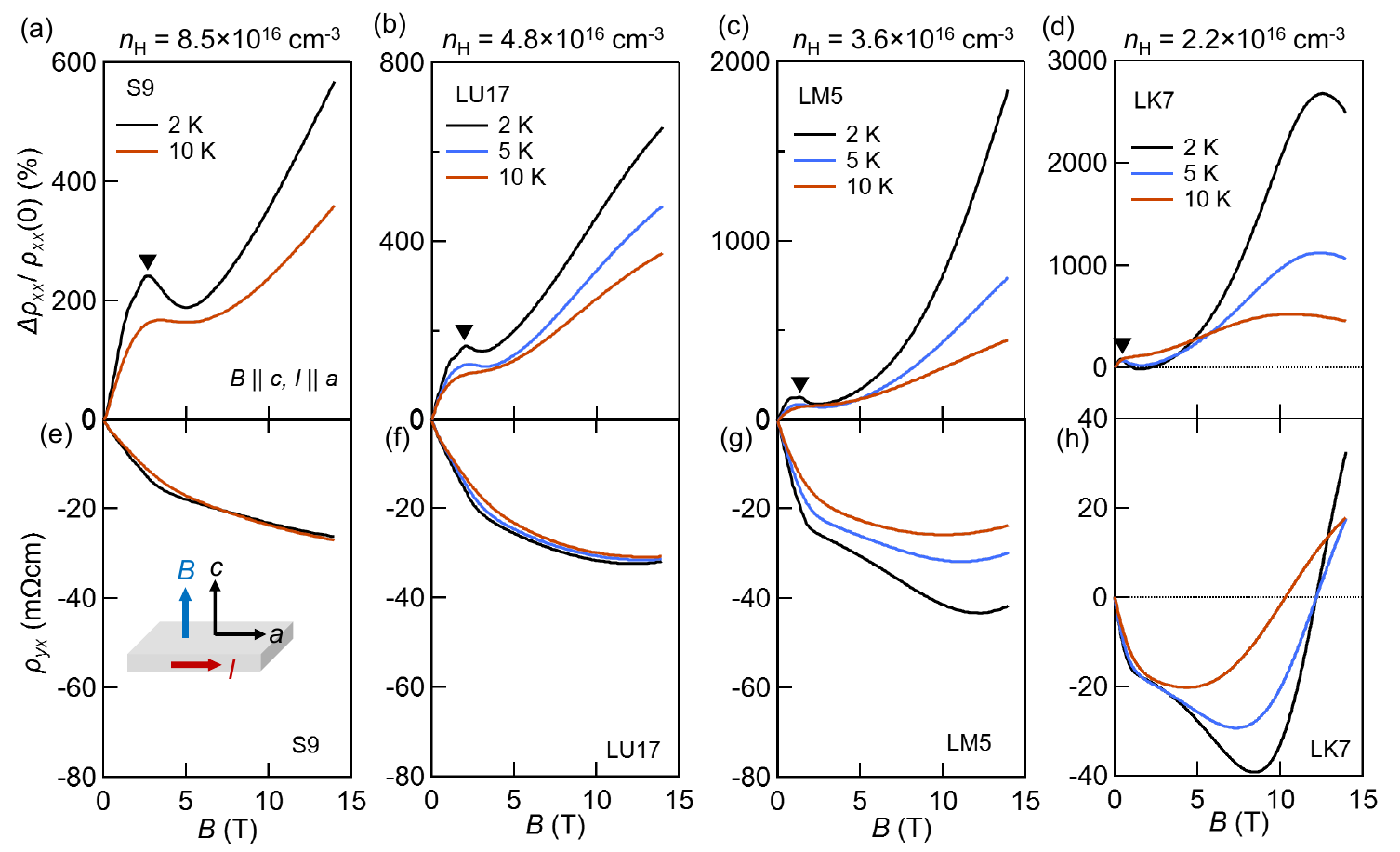}
\caption{\textbf{Comparison of magnetotransport of samples with different filling.}  
MR ratio of (a) sample S9 ($n_\mathrm{H}=8.5\times10^{16}\,$cm$^{-3}$), (b) LU17 ($n_\mathrm{H}=4.8\times10^{16}\,$cm$^{-3}$), (c) LM5 ($n_\mathrm{H}=3.6\times10^{16}\,$cm$^{-3}$), and (d) LK7 ($n_\mathrm{H}=2.2\times10^{16}\,$cm$^{-3}$) at several temperatures. The black triangle indicates the magnetic field where a kink structure appears in the MR.
Hall resistivity of (e) sample S9, (f) LU17, (g) LM5, and (h) LK7.
}\label{MR_Hall_FourSamples}
\end{figure}

Figures \ref{MR_Hall_FourSamples}(a)-(d) show the magnetoresistance at several temperatures. The MR ratio at a high magnetic field is commonly enhanced with decreasing temperature for all the samples. The peak of MR around $2 \, \mathrm{T}$ disappears above $10 \, \mathrm{K}$. We also observe a systematic change in Hall resistivity ($\rho_{yx}$) with {\nH}, as shown in Figs. \ref{MR_Hall_FourSamples} (e)-(h). $\rho_{yx}$ shows a linear field dependence below the magnetic field where a kink appears in the MR and $\rho_{yx}$. At high fields, $\rho_{yx}$ shows a dip structure around $10 \, \mathrm{T}$ in the sample LU17, which becomes significant and shifts to lower magnetic fields with decreasing {\nH}. In the sample LK7 with the lowest {\nH}, $\rho_{yx}$ shows a sign change at $12 \, \mathrm{T}$, where a peak appears in the MR.

\begin{figure}[h]%
\centering
\includegraphics[trim=0cm 0.1cm 0.0cm 0.0cm,clip,width=0.7\textwidth]{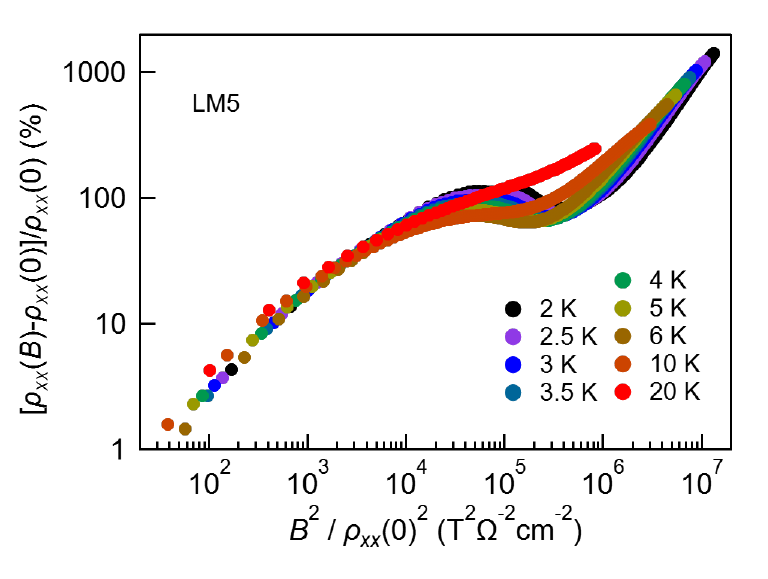}
\caption{\textbf{Kohler's plot of a representative sample LM5.}  
MR ratio plotted as a function of $B^2/${\Rhoxx}$(0)^2$. At low magnetic fields, MR at different temperatures falls onto a single curve, following Kohler's rule~\cite{1989ABPippardCambridge}.
}\label{Kohler}
\end{figure}

In many conventional nonmagnetic metals or semiconductors, the MR ratio is often scaled to $(B \tau)^2$, which is known as Kohler's rule \cite{1989ABPippardCambridge}. Here, we plotted the MR ratio of the sample LM5 as a function of $B^2/\rho_{xx}(0)^2$ in Fig. \ref{Kohler}. The data at various temperatures nearly fall onto a single line at low fields in a log-log plot, in agreement with Kohler's rule. This linear behavior in a log-log plot indicates that the MR ratio scales with $B^2$ at low fields. In contrast, under higher magnetic fields, where the SdH oscillations appear, the MR ratio curves no longer fall onto the same curve. This breakdown of Kohler's rule indicates that the magnetotransport under high magnetic fields cannot be described by a single relaxation time in a semiclassical picture, as suggested by the emergence of SdH oscillations (see below).

\subsection{Shubnikov-de Haas oscillations.} 
\begin{figure}[h]%
\centering
\includegraphics[trim=0cm 0.1cm 0.0cm 0.0cm,clip,width=0.95\textwidth]{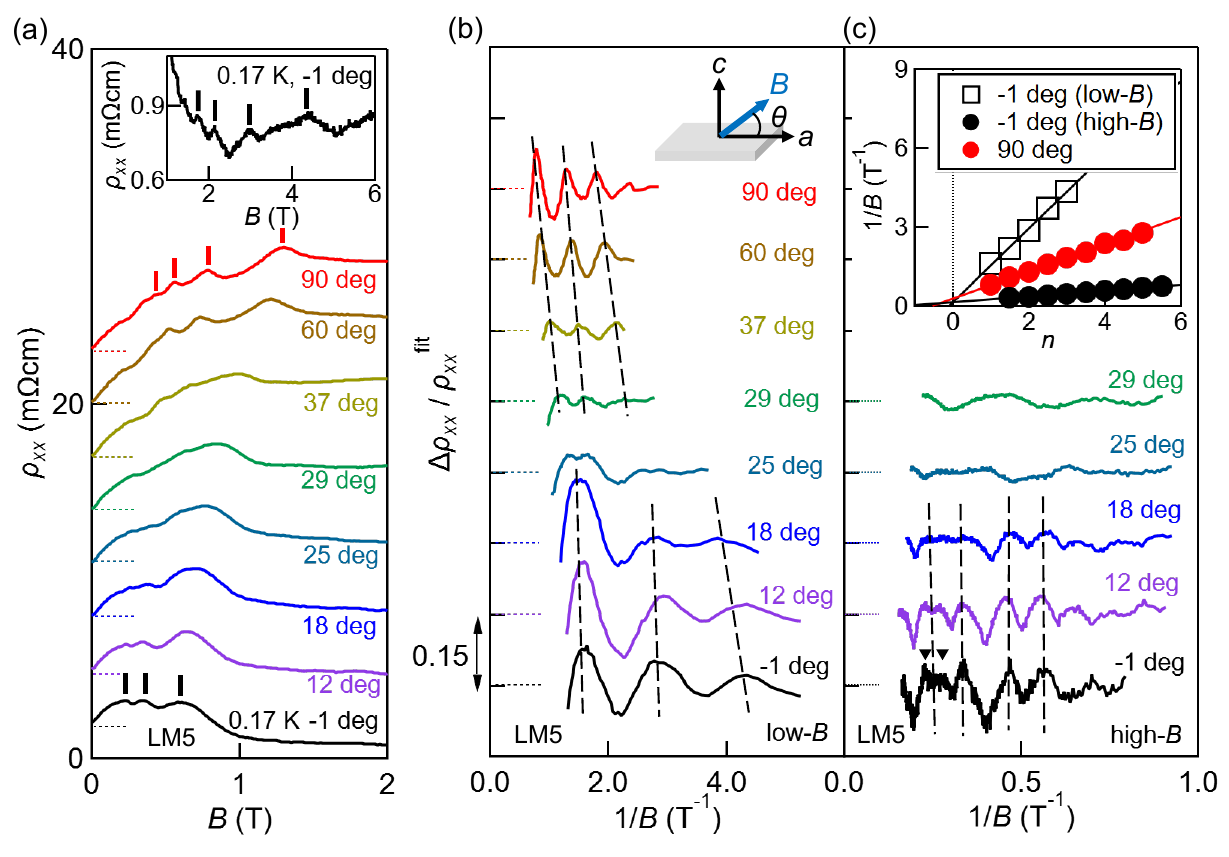}
\caption{\textbf{Angular dependence of Shubnikov-de Haas oscillations.}  (a) Angular dependence of the magnetoresistivity at 0.17 K with offset for clarity. The magnetic field tilted in $a$-$c$ plane and $\theta$ is defined as the tilting angle measured from $a$-axis. The inset shows the magnified view of the magnetoresistivity for $\theta$=-1 deg. The vertical lines indicate the position of the peak of SdH oscillations.
(b) Angular dependence of the SdH oscillations in the low-$B$ regime below 2 T with offset.
(c) Angular dependence of the SdH oscillations in the high-field regime above 2 T with offset. The inset shows the Landau index plot of SdH oscillations in the low-$B$ regime for {\Bparaa} (black), those in the high-field regime for {\Bparaa} (blue), and those for {\Bparac} (orange). The dotted lines in (b) and (c) are guides for the eye and triangles in (c) denote the position of split two peaks of $n$=1 Landau level.
}\label{SdH_AngleDep}
\end{figure}

To clarify the fundamental transport parameters, we analyze SdH oscillations under various temperatures and different field directions. Here, we focus on the results of sample LM5 ($n_\mathrm{H} = 3.6 \times10^{16}\,$cm$^{-3}$) as a typical example. The magnetic field is tilted within the $ac$-plane, and the angle between the field and $a$-axis is defined as $\theta$ [see the inset to Fig.~\ref{SdH_AngleDep}(b)]. Figure~\ref{SdH_AngleDep}(a) shows the angular dependence of the magnetoresistivity at $0.17 \, \mathrm{K}$. When the magnetic field is applied along the $c$-axis ($\theta = 90 \, \mathrm{deg}$), the MR gradually increases, accompanying the beating pattern due to the SdH oscillations below $2 \, \mathrm{T}$. 
When $\theta$ is close to $0 \, \mathrm{deg}$, clear SdH oscillations appear not only in the low-field regime below $1 \, \mathrm{T}$ but also in the relatively high-field regime above $2 \, \mathrm{T}$ [see the inset to Fig. \ref{SdH_AngleDep}(a)]. 

To analyze the SdH oscillations quantitatively, we subtract the non-oscillating background ({\RhoxxFit}) from the raw data of the MR, and then normalize the oscillating component ({\DeltaRhoxx}) by {\RhoxxFit}. The oscillatory component ({\DeltaRhoxx}/{\RhoxxFit}) is plotted against the inverse of the magnetic field in Figs. \ref{SdH_AngleDep}(b) and (c). We first plot the Landau level (LL) fan-diagram of the SdH oscillations to extract the oscillation frequency {\BF} and phase shift $\phi$ as shown in the inset to Fig. \ref{SdH_AngleDep}(c), following the Lifshitz-Onsager quantization rule $B_{\rm F}/B = n - \phi$ with $n$ being the Landau index. Here, we assigned the peak and dip positions to integer and half-integer in the LL fan diagram, respectively. The effect of Zeeman splitting is likely to be negligibly small for SdH oscillations in the low-$B$ regime for {\Bparaa} and those for {\Bparac}, judging from the fact that the LL fan diagram shows a nearly linear behavior. On the other hand, the $n=1$ peak in the high-field regime for {\Bparaa} is split into two [see Fig. \ref{SdH_AngleDep}(c)]. We estimate $g$-factor as $3.5\pm1.5$ using the size of the splitting of the $n=1$ peak [see Fig. \ref{FigS_IndexPlot}~\cite{Supple}]. Although the uncertainty of the $g$-factor is not negligibly small, it is unlikely to affect the following analysis of the filling dependence of SdH oscillations and magnetoresistance.

For the analysis of the LL fan-diagram in the high-field regime for {\Bparaa}, we use the data with the valley of $n>$1.5, where the Zeeman splitting is negligible. The extracted {\BF} in the low-$B$ for {\Bparaa}, that in the high-$B$ for {\Bparaa}, and that for {\Bparac} are $0.71 \, \mathrm{T}$, $9.2 \, \mathrm{T}$, and $1.9 \, \mathrm{T}$, respectively. The extracted $\phi$ for those oscillations are -0.1, -1.1, and -0.5, respectively. Considering the LL fan diagram and the absence of oscillations in the high field regime, we conclude that these Fermi pockets reach the quantum limit at $0.6 \, \mathrm{T}$, $4.2 \, \mathrm{T}$, and $1.3 \, \mathrm{T}$, respectively.

\begin{figure}[h]%
\centering
\includegraphics[trim=0cm 0.1cm 0.0cm 0.0cm,clip,width=0.95\textwidth]{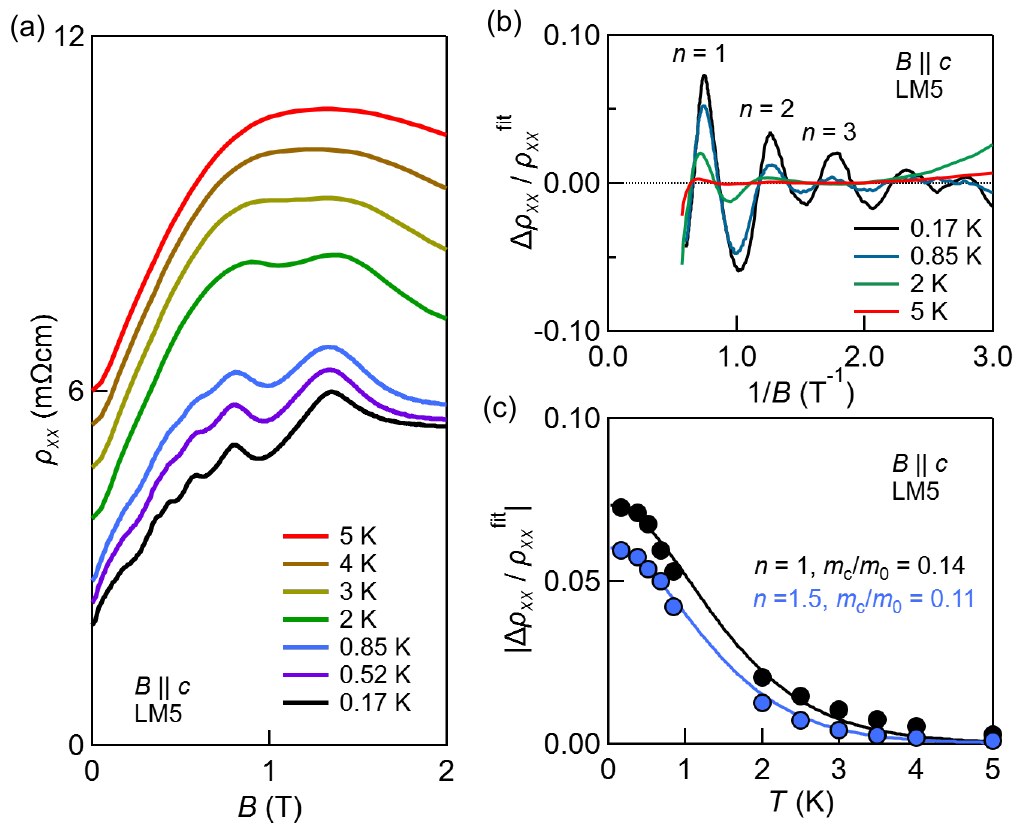}
\caption{\textbf{Mass analysis on SdH oscillations for {\Bparac}.}
(a) Temperature dependence of magnetoresistivity for {\Bparac}. 
(b) Oscillatory component of SdH oscillations in the low-$B$ regime for {\Bparac}.
(c) Temperature dependence of the oscillatory amplitude at the Landau indices of $n$=1 and $n$=1.5. The effective mass is extracted at each Landau index by fitting the experimental data with Eq.~(\ref{equ_LKformular}). The fitting result is shown as a solid line.
}\label{MassAnalysis_Bc}
\end{figure}

\begin{figure}[h]%
\centering
\includegraphics[trim=0cm 0.1cm 0.0cm 0.0cm,clip,width=0.9\textwidth]{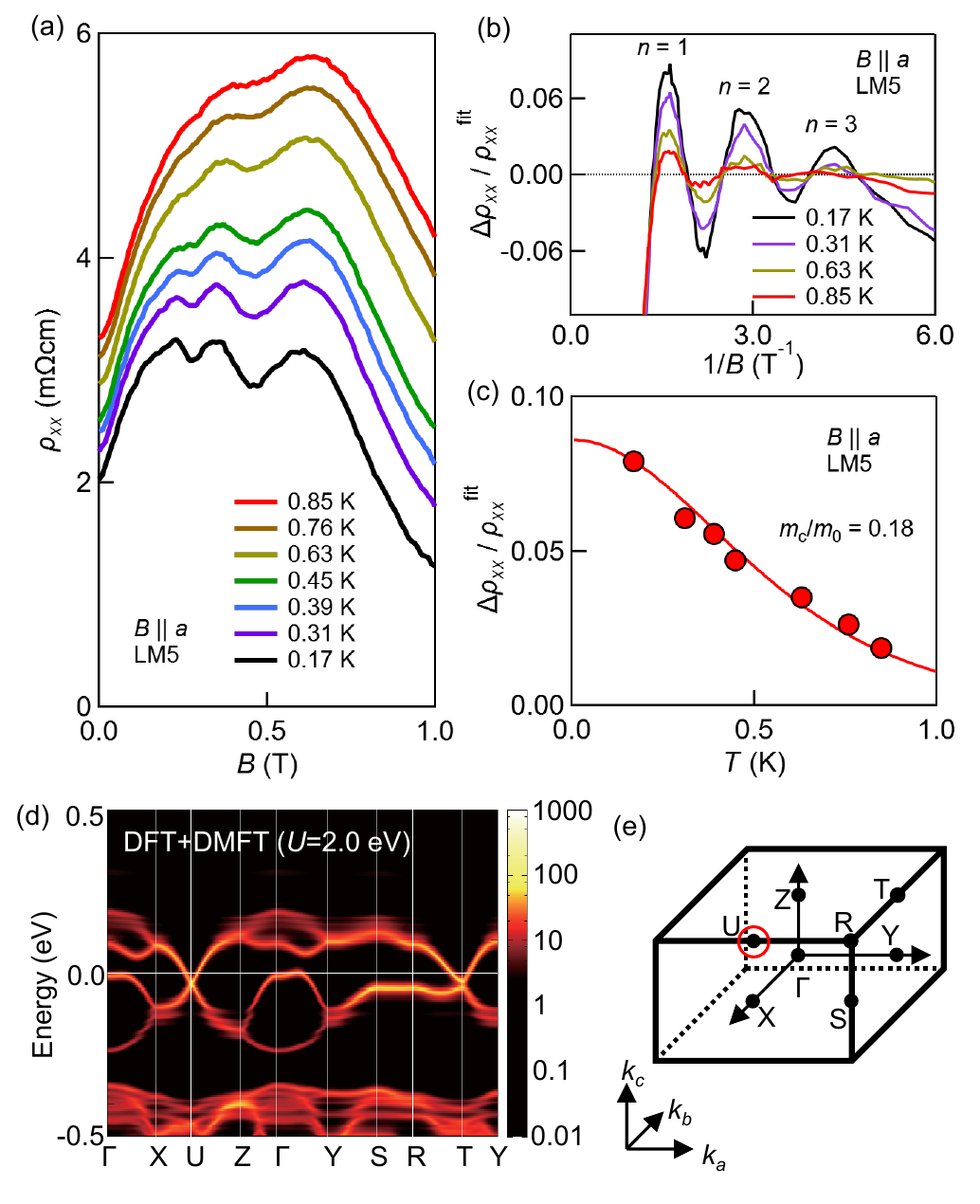}
\caption{\textbf{Mass analysis on SdH oscillations for {\Bparaa}.}
(a) Temperature dependence of magnetoresistivity for {\Bparaa}. 
(b) Oscillatory component of SdH oscillations in the low-$B$ regime for {\Bparaa}.
(c) Temperature dependence of the oscillatory amplitude at the Landau indices of $n$=1 and a fitting result to extract the effective mass (dotted line). 
(d) Band structure calculated by density functional theory (DFT) and dynamical mean field theory (DMFT) with the Hubbard $U_\mathrm{eff} = 2.0$ eV, which is reproduced from Ref.~\cite{2019FujiokaNatCommun}. The scale bar denotes the magnitude of the spectral function.
(e) Position of Dirac line node (red line) in Brillouin zone.
}\label{MassAnalysis_Ba}
\end{figure}

To further extract transport parameters such as the effective mass {\massc} and Fermi velocity {\vF}, we analyze the temperature dependence of the SdH oscillations. Figure \ref{MassAnalysis_Bc}(a) shows the MR for {\Bparac} at several temperatures. The SdH oscillations are suppressed with increasing temperature [Fig. \ref{MassAnalysis_Bc}(b)]. From the thermal damping of the oscillation amplitude, we derived the cyclotron mass $m_{\rm c}$ to be $0.11 \, m_{0}$ [see Fig. \ref{MassAnalysis_Bc}(c)]. Here, the amplitude of the dip for $n=1.5$ is used for the analysis of the oscillation of an electron pocket at the U-point. We could not obtain a good fit to the temperature dependence of the $n=1$ peak, possibly due to the breakdown of the Lifshitz-Kosevich formula under a high magnetic field. The Fermi velocity {\vF} is estimated to be $7.9 \times 10^4 \, \mathrm{m/s}$ using the relation $\hbar k_{\rm F} = m_{\rm c} v_{\rm F}$. Here, $\hbar$ and $k_{\rm F}$ are Planck's constant divided by $2\pi$ and Fermi wave number, respectively. We also perform the analysis for the temperature dependence of the SdH oscillations observed in the low-$B$ for {\Bparaa} [see Fig. \ref{MassAnalysis_Ba}(a)]. From the thermal damping of the oscillation amplitude of $n=1$, {\massc} and {\vF} are extracted to be $0.18 \, m_{0}$ and $3.0 \times 10^4 \, \mathrm{m/s}$, respectively [see Fig. \ref{MassAnalysis_Ba}(c)]. 

Having established the transport parameters from the analysis of SdH oscillations, we discuss the assignment of each oscillatory frequency. The theoretical calculation of electronic structure indicates the presence of electron pockets at the U-point and at the T-point, and a hole pocket at the $\Gamma$-point in the Brillouin zone [see Fig.~\ref{MassAnalysis_Ba}(d) and (e)]. In the previous study, the SdH oscillations in the low-$B$ for {\Bparac} and those in the high-$B$ for {\Bparaa} are assigned to the electron pockets near the Dirac line node \cite{2019FujiokaNatCommun}. In addition, the SdH oscillations from the hole pocket at the $\Gamma$-point with a relatively large electron mass have been observed in the high field regime above $20 \, \mathrm{T}$ \cite{2022RYamadanpj}. Therefore, the newly observed SdH oscillations in the low-$B$ for {\Bparaa} likely originate from the electron pocket at the T-point. 
The smaller {\vF} for the oscillations in the low-$B$ for {\Bparaa} (electron pocket at the T-point), as compared to those for {\Bparac} (electron pocket at the U-point), supports our assignment, since the band calculation predicts that the former is a quadratic band and the latter is a gapless Dirac-like linear band as shown in Fig.~\ref{MassAnalysis_Ba}(d)~\cite{2012JMCarterPRB, 2012MAZebPRB, 2013HZhangPRL}. 

\subsection{Filling dependence of transport parameters of Dirac electrons.}
\begin{figure}[h]%
\centering
\includegraphics[trim=0cm 0.1cm 0.0cm 0.0cm,clip,width=0.75\textwidth]{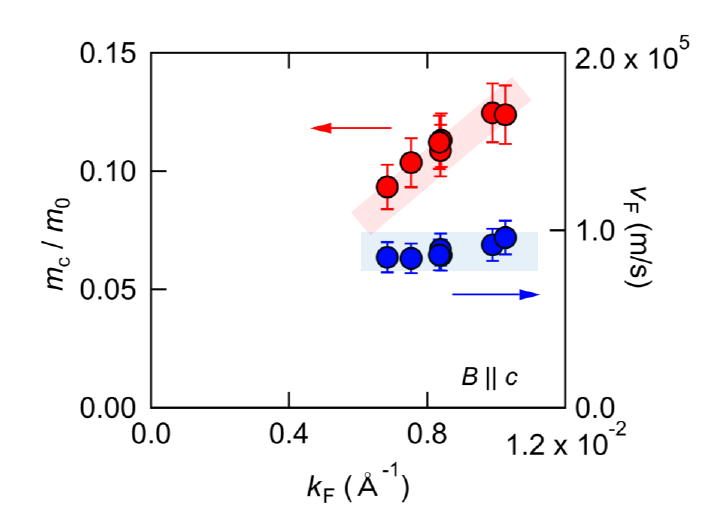}
\caption{\textbf{Filling dependence of transport parameters of Dirac electrons.}
Effective mass \mass{} and Fermi velocity \vF{} plotted against Fermi wave number \kF{}, estimated from SdH oscillations. While \mass{} proportionally increases with \kF{}, \vF{} is independent of the filling, suggesting that the band dispersion is linear.
}\label{kFDep_mass}
\end{figure}

On the basis of the extracted transport parameters, we discuss the filling dependence of the transport parameters of Dirac electrons among the samples with different band fillings. Figure \ref{kFDep_mass} shows the {\kF} dependence of {\massc} and {\vF}. The cyclotron mass {\massc} decreases almost in proportion to {\kF}, while the Fermi velocity is nearly independent of the band filling. This result suggests that the electron pockets originate from the electronic state hosting the $k$-linear dispersion, which is consistent with the theoretical calculation and photoemission spectroscopy~\cite{2012MAZebPRB, 2012JMCarterPRB, 2015NiePRL}. This is because the cyclotron mass is expressed as $m^* = \hbar k_{\mathrm{F}} / v_{\mathrm{F}}$ due to momentum conservation, while $v_{\mathrm{F}}$ stays constant independent of the band filling for linear dispersion. We stress that extremely small energy scales can be resolved by utilizing clear SdH oscillations. The range of the Fermi energy of the samples shown in Fig. \ref{kFDep_mass} is $3.8 \sim 6.5 \,$meV, which is difficult to detect in the standard photoemission spectroscopy. 

Next, we discuss the filling dependence of the carrier mobility of Dirac electrons. We assume that the contribution from the hole pocket at the $\Gamma$-point and the electron pocket at the T-point is likely small because the former has a large effective mass ($1.8 \,${\massz}~\cite{2022RYamadanpj}) and the latter has a small Fermi surface, as compared to the Dirac electron pocket at the U-point. This assumption is supported by the proportionality between the carrier density of Dirac electrons from SdH oscillations (\nthreeD{}) and the carrier density from the Hall effect using a single carrier model (\nH{}) in Fig.~\ref{SFig_kF_nH}(b)~\cite{Supple}; $n_\mathrm{3D}$ is estimated to be approximately $0.7 \, n_\mathrm{H}$, suggesting that \nH{} is roughly determined by Dirac electron pockets around U-point. In the following, we focus on the results for $B \parallel c$. 

\begin{figure}[h]%
\centering
\includegraphics[trim=0cm 0.1cm 0.0cm 0.0cm,clip,width=0.9\textwidth]{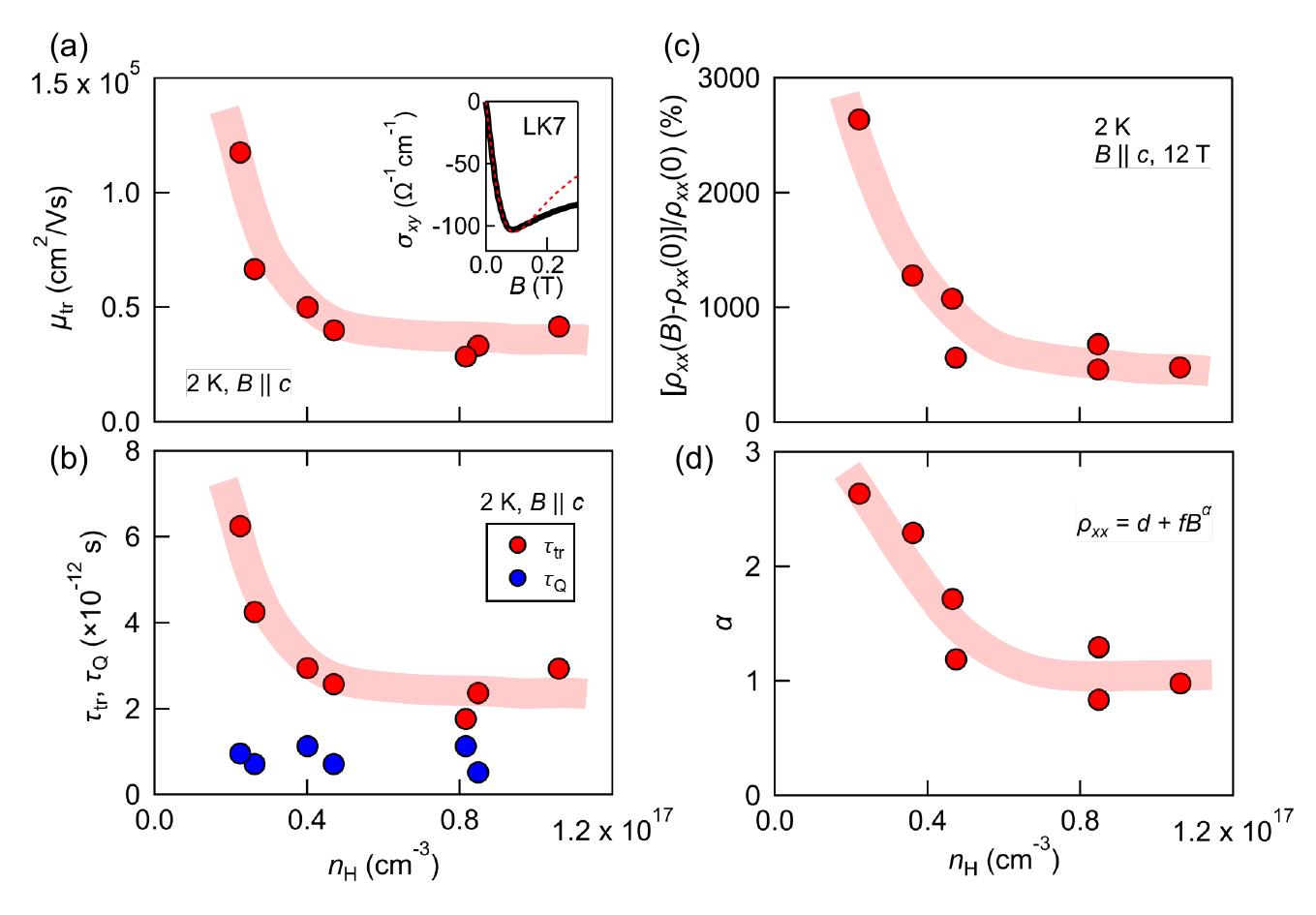}
\caption{\textbf{Filling dependence of carrier mobility and MR.}
Carrier density {\nH} dependence of (a) Transport mobility {\mutr}, (b) transport scattering time {\tautr} and quantum relaxation time {\tauQ}, (c) MR ratio at $14 \, \mathrm{T}$ for $B \parallel c$, and (d) the exponent of MR $\alpha$ at high fields using the fitting function $\rho_{xx} = d + f B^{\alpha}$. Light red lines are guides to the eye. The inset to panel (a) shows the magnified view of the Hall conductivity at low fields with a peak at $B_\mathrm{peak}$ (black triangle). The carrier mobility \mutr{} is defined as $\mu_\mathrm{tr} = 1/B_\mathrm{peak}$. The low-field behavior of the Hall conductivity around  $B_\mathrm{peak}$ is reproduced by a single-carrier Drude model (red dotted line).
}\label{nHDep_MR}
\end{figure}

We estimate the carrier mobility of Dirac electrons from the Hall conductivity~\cite{2014TLiangNatMater, 2019FujiokaNatCommun}. The Hall conductivity {\Gxy} is calculated as $\rho_{yx}/({\rho_{xx}}^2+{\rho_{yx}}^2)$, where $\rho_{xx}$ and $\rho_{yx}$ are the longitudinal and Hall resistivity, respectively. We determine the transport mobility {\mutr} as the inverse of the peak field of the Hall conductivity ($B_\mathrm{peak}$), i.e., $\mu_\mathrm{tr} = 1/B_\mathrm{peak}$ without performing a fitting. The validity of this analysis is confirmed by reproducing the low-field behavior of the Hall conductivity by a semiclassical Drude model $\sigma_{xy} =  n_{\rm tr} e {\mu}_{\rm tr} \frac{{\mu}_{\rm tr} B}{1 + ({\mu}_{\rm tr} B)^2}$  [see the inset to Fig. \ref{nHDep_MR}(a)]. We plot {\mutr} against the Fermi wave number ({\kF}) to investigate the band filling dependence. Here, $k_\mathrm{F}$ is extracted from the quantum oscillations, assuming an isotropic Fermi surface, i.e., $S_\mathrm{F} = \pi k_\mathrm{F}^2$, where $S_\mathrm{F}$ is the extremal cross-sectional area of the Fermi surface. The mobility {\mutr} increases with decreasing {\kF} and reaches $1.2 \times 10^5$ {\muUnit}, which is the highest level among the known bulk oxide semiconductors, to the best of our knowledge \cite{2019FujiokaNatCommun, 2021PRMNPQuirk}.

We further estimate the transport scattering time {\tautr} and quantum scattering time {\tauQ} in Fig. \ref{nHDep_MR}(b). {\tautr} is extracted from the mobility {\mutr} using the relation $\mu_{\rm tr} = e v_{\rm F} \tau_{\rm tr}/\hbar k_{\rm F}$, while {\tauQ} is obtained from the Dingle plot of SdH oscillations [see Fig.~\ref{FigS_tauQ}~\cite{Supple}]. As shown in Fig. \ref{nHDep_MR}(b), {\tautr} is enhanced with decreasing {\kF}, while {\tauQ} remains nearly constant. In general, {\tauQ} is determined by all scattering processes that cause the Landau level broadening. On the other hand, {\tautr} is mainly determined by the backward scattering, which accompanies the large variation of momentum. Therefore, the enhancement of {\tautr} may indicate the significant suppression of the background scattering in the low-carrier density regime of the Dirac cone.


\subsection{Origin of large transverse magnetoresistance (MR).}
In the following, we discuss the origin of the giant transverse MR for {\Bparac} considering the carrier density {\nH} dependence. 
The {\nH} dependence of the MR ratio at $2 \, \mathrm{K}$ and $12 \, \mathrm{T}$ is shown in Fig. \ref{nHDep_MR}(c). 
The MR ratio is enhanced in the low-{\nH} regime; in particular, it exceeds 2,000 \% for the sample with the lowest carrier density ({\nH}$ = 2.2 \times 10^{16} \,${\nUnit}). 
To further elucidate the origin of the MR of {\CaIrO}, we extract the exponent of the MR with respect to the magnetic field. We fit the MR with the formula $\rho_{xx} = d + f B^{\alpha}$ at high magnetic fields above $5\,\mathrm{T}$. Here, $d$, $f$, and $\alpha$ are fitting parameters independent of the magnetic field. We plot $\alpha$ as a function of {\nH} in Fig.~\ref{nHDep_MR}(d). While $\alpha$ is close to 1 when {\nH} is large, it is enhanced in the low-{\nH} regime and exceeds 2 at {\nH} $< 4 \times 10^{16} \,${\nUnit}. Such variation of $\alpha$ suggests that the transport scattering mechanism changes in different regimes of {\nH}. 

The $B$-linear MR, which appears in the high-{\nH} regime of {\CaIrO}, has been reported in various topological semimetals~\cite{2015JXiongSience, 2014TLiangNatMater}. Several theories have been proposed to explain $B$-linear MR under high magnetic fields, including a guiding center model and Abrikosov's model~\cite{2015JCWSongPRB, 1998AAAbrikosovPRB}. Both models consider non-charged impurities due to the screening of Coulomb interactions under the high magnetic field, which may also be applicable in the high-{\nH} regime of {\CaIrO}. 

However, when {\nH} becomes sufficiently small, the screening by conduction electrons is expected to be suppressed, and the long-range Coulomb interaction should play an important role. It has been theoretically pointed out that the transport property of topological semimetals varies largely depending on the type of impurities: i.e., short-range or charged (Coulomb) impurities~\cite{2015SDSarma, 2015JKlierPRB}. In \CaIrO{}, the Thomas-Fermi screening length $\lambda_\mathrm{TF}=\sqrt{\epsilon/e^2D(E_\mathrm{F})}$ is on the order of 1 to $10 \,$nm, i.e., slightly larger than the lattice constant, using the dielectric constant $\epsilon \sim 100 \, \epsilon_0$ at zero field~\cite{2021JFujiokaPRB} and the density of states $D(E_\mathrm{F})$ estimated from the quantum oscillation analysis. Although it is difficult to estimate the precise value of $\lambda_\mathrm{TF}$, the relatively large $\lambda_\mathrm{TF}$ supports our scenario based on the variation of the type of impurities with band fillings. 

In this sense, the enhanced effect of Coulomb impurities may explain the change of MR from the $B$-linear dependence to the $B$-nonlinear behavior with the exponent $\alpha$ exceeding 2 with decreasing {\nH} in the quantum limit of Dirac electrons. It is known that the MR follows the square of the magnetic field for parabolic bands~\cite{1988AbrikosovBook}; however, it is pointed out that the exponent of the MR can largely depend on the type of impurities and the strength of the magnetic field for linear bands due to a significant broadening of Landau levels~\cite{2015JKlierPRB}. Specifically, under the assumption of the charged impurity and low temperature, the exponent of MR is predicted to be 2 for magnetic fields smaller than the threshold determined by impurity strength $\sim N_\mathrm{imp}^{-1/3}v$, where $N_\mathrm{imp}$ and $v$ are the density of charged impurities and Fermi velocity, respectively. A more detailed comparison with theory requires an estimate of the strength of the impurity potential. We note that a similar variation of MR is reported in {\CdAs}~\cite{2014TLiangNatMater}. In case of {\CdAs}, while a $B$-linear MR is observed in polycrystals and single crystals with lower mobility ($1 \times 10^{5}$ to $1.5 \times 10^{5} \,${\muUnit}), the MR evolves to the $B^2$-profile for single crystals with higher mobility ($1.5 \times 10^{5}$ to $1 \times 10^{7} \,${\muUnit}).

In other correlated topological semimetals, giant MR has been observed. It has been argued that the reconstruction of the electronic state upon the field-induced Mott transition causes the giant negative MR in pyrochlore iridates~\cite{2015KUedaPRL, 2015ZTianNatPhys}. In contrast, a large positive $B$-linear MR due to the high-mobility electrons is also observed in the quantum limit of Dirac electrons in SrNbO$_3$~\cite{2021SciAdvJMOk}. The latter is partly in common with the behavior of \CaIrO{}, whereas the effect of charged impurities is likely screened by trivial bands with carrier density of $\sim10^{22} \, \mathrm{cm}^{-3}$ in SrNbO$_3$. In {\CaIrO}, high-mobility Dirac electrons coexist with a semimetallic state with a small density of states, where the screening of Coulomb interaction is suppressed in the QL. Such a semimetallic state with dilute carrier density in the strongly correlated regime may cause the unique giant MR in \CaIrO{}.

\section{Conclusions}
In summary, we have investigated the carrier density dependence of transverse magnetoresistance in the correlated Dirac semimetal {\CaIrO}. Both the mobility and transverse MR ratio are enhanced with decreasing the carrier density and exceed $1.0 \times 10^5 \,${\muUnit} ($2 \, \mathrm{K}$) and $2,000 \,$\% ($2 \, \mathrm{K}$ and $12 \, \mathrm{T}$), respectively. The analysis of Shubnikov-de Haas oscillations and Hall conductivity shows that the electrons nearby the Dirac line node govern the charge transport at low temperatures. In particular, the Fermi velocity of the carriers hardly changes as the carrier density changes, suggesting that the energy dispersion close to the line node is approximately described by the $k$-linear dispersion. Since the relaxation time increases with decreasing carrier density, it is likely that the mobility enhancement in the dilute carrier density is attributed to both the $k$-linear band dispersion, or equivalently, the reduced effective transport mass, and the enhancement of relaxation time. Moreover, we found that the MR ratio is of $B$-linear type in the higher carrier density region, whereas it shows $B^{\alpha}$ ($\alpha > 2$) behavior in the low carrier density region. The former may be explained by the electron scattering by the non-charged impurities, as often seen in conventional Dirac semimetals. However, in the low-{\nH} regime where screening by conduction electrons is expected to be suppressed, the long-range Coulomb interaction may lead to the $B$-nonlinear and large MR.

\newpage

\textbf{Acknowledgement}\\
Authors thank A. Tsukazaki, M. Kawasaki, N. Nagaosa, M. Tokunaga, S. Ishiwata, A. Yamamoto, M. Hirschberger, M. Masuko, and R. Kaneko for fruitful discussion. This work was supported by a Japan Society for the Promotion of Science KAKENHI (Grants No. 20J21312, No. 16H00981, No. 18H01171, No. 18H04214, No. 16H06345, No. 23H05431, No. 22K20348, No. 24H01604, and 25K17336) from the MEXT and by PRESTO (Grant No. JPMJPR15R5 and JPMJPR259A) and CREST (Grants No. JPMJCR16F1), Japan Science and Technology Japan, and TRIP Initiative program (Many-body physics).

\clearpage
\begin{center}
\Large{Main text Figures}
\end{center}

\clearpage

\clearpage

\clearpage

\clearpage

\clearpage

\clearpage
\begin{table}[h]
	\centering
		\begin{tabular}{|c|c|c|c|}\hline
			\ Sample name \ & \ {\nthreeD} ({\nUnit}) \ & \ \ {\nH} ({\nUnit}) \ \ & \ {\mutr} ({\muUnit}) \ \\ \hline \hline
			S9 & $6.5 \times 10^{16}$ & $8.5 \times 10^{16}$ & $3.3 \times 10^{4}$  \\ \hline
			S5 & $7.3 \times 10^{16}$ & $1.1 \times 10^{17}$ & $5.0 \times 10^{4}$  \\ \hline
			LU17 & $4.0 \times 10^{16}$ & $4.8 \times 10^{16}$ & $4.0 \times 10^{4}$ \\ \hline
			LU26 & $4.0 \times 10^{16}$ & $8.5 \times 10^{16}$ & $2.9 \times 10^{4}$ \\ \hline
			LM5 & $2.9 \times 10^{16}$ & $3.6 \times 10^{16}$ & $5.0 \times 10^{4}$ \\ \hline
			LM16 & $3.9 \times 10^{16}$ & $4.7 \times 10^{16}$ & $6.7 \times 10^{4}$ \\ \hline
			LK7 & $2.2 \times 10^{16}$ & $2.2 \times 10^{16}$ & $1.2 \times 10^{5}$ \\ \hline
		\end{tabular}
	\caption{Sample list summarizing the carrier density and mobility. {\nthreeD} is the carrier density extracted from SdH oscillations, defined as $n_\mathrm{3D} = d_\mathrm{s}d_\mathrm{g}k_\mathrm{F}^3 / 6\pi^2$. Here, $d_\mathrm{s} = 2$ and $d_\mathrm{g} = 2$ are assumed for the spin and orbital degeneracies, respectively, based on the Landau level calculations~\cite{2022RYamadanpj}. \nH{} is extracted from the Hall resistivity at zero field using a single-carrier model. {\mutr} is the carrier mobility defined as the inverse of the magnetic field where the Hall conductivity shows a peak ($B_\mathrm{peak}$), i.e. $\mu_\mathrm{tr} = 1/B_\mathrm{peak}$.
    }
    \label{Table_SampleList}
\end{table}

\clearpage

\clearpage

\end{document}